\documentclass[12pt]{article}
\usepackage{amsmath}
\usepackage{amssymb}
\usepackage{amsfonts}
\newcommand{\ba}{\begin{array}{l}}
\newcommand{\ea}{\end{array}}
\newcommand{\beq}{\begin{equation}}
\newcommand{\eeq}{\end{equation}}
\newcommand{\bea}{\begin{eqnarray}}
\newcommand{\eea}{\end{eqnarray}}
\usepackage{epsfig}
\usepackage{graphicx}
\usepackage{color}

\definecolor{dyellow}{rgb}{1.,0.8,.0}
\definecolor{myblue}{rgb}{.1,.1,.7}
\definecolor{dcyan}{rgb}{.0,.6,.6}
\definecolor{dmagenta}{rgb}{0.6,0.0,0.6}
\definecolor{brown}{rgb}{0.6,0.2,0.}
\definecolor{darkblue}{rgb}{.0,.0,0.5}
\definecolor{darkred}{rgb}{0.75,0.0,0.0}
\definecolor{orange}{rgb}{1.,.6,.0}
\definecolor{dorange}{rgb}{0.8,.4,.0}
\definecolor{darkgreen}{rgb}{0.0,0.6,0.0}
\definecolor{purple}{rgb}{.4,.0,.4}

\def\bc{\begin{center}}
\def\ec{\end{center}}
\def\be{\begin{eqnarray}}
\def\ee{\end{eqnarray}}

\newcommand{\omits}[1]{}

\begin{document}
\begin{center}
{\Large \bf { A possible scenario of the Pioneer anomaly in the framework of Finsler geometry}}\\
  \vspace*{1cm}
Xin Li$^{\ast,\ddagger}$ \footnote{lixin@itp.ac.cn} and Zhe Chang$^{\dagger,\ddagger}$ \footnote{changz@mail.ihep.ac.cn}\\
\vspace*{0.2cm} {\small $^\ast$Institute of Theoretical Physics,
Chinese Academy of Sciences, 100190 Beijing, China}\\
{\small $^\dagger$Institute of High Energy Physics, Chinese Academy
of Sciences, 100049 Beijing, China}\\
{\small $^\ddagger$Theoretical Physics Center for Science Facilities, Chinese Academy of Sciences}\\

\bigskip

\end{center}
\vspace*{2.5cm}

%
\begin{abstract}\baselineskip=30pt
The weak field approximation of geodesics in Randers-Finsler
space is investigated. We show that a Finsler structure of
Randers space corresponds to the constant and sunward anomalous
acceleration demonstrated by the Pioneer 10
and 11 data. The additional term in the geodesic equation acts as
``electric force'', which provides the anomalous acceleration.
 \vspace{1cm}
\begin{flushleft}
\end{flushleft}
\end{abstract}

\newpage
\baselineskip=30pt

Newton's theory of gravitation was proposed almost three hundred and fifty years
ago. Einstein's general relativity reveals the intrinsic geometric property of gravity. Newton's
theory of gravitation is the main guideline of the celestial
mechanics, especially for the solar system. General relativity
provides small corrections. It is well-known that the Newton and Einstein's theory of gravitation still faces problems. One of them is that the flat rotation curves of spiral galaxies violate the prediction
of Newton and Einstein's gravity. Another is related with recent astronomical
observations\cite{Riess}. Our universe is acceleratedly expanding.
This result can not be obtained directly from Einstein's gravity and
his cosmological principle. In fact, new puzzle has also arisen in the solar
system. That is the Pioneer anomaly. The Pioneer spacecrafts are excellent tools for dynamical
astronomy studies in the solar system. The radio metric data from the
Pioneer 10/11 spacecraft indicate the presence of a small,
anomalous, Doppler frequency drift over the range of
20--70 astronomical units\cite{Anderson}. The drift is blue-shift,
uniformly changing with a rate of $~6\times10^{-9}
Hz/s$\cite{Turyshev1}. It has revealed an anomalous constant sunward
acceleration, $a_P =
(8.74\pm1.33)\times10^{-10} m/s^2$.

The Pioneer data have been studied in three different navigational
computer programs. Namely: the JPL's Orbit Determination Program
(ODP), the Aerospace Corporation's CHASPM code extended for deep
space navigation\cite{Anderson}, and a code written in the Goddard
Space Flight Center\cite{Markwardt}. These data analysis all confirm
the existence of anomalous acceleration with the following basic
properties: the direction of the anomalous acceleration of the
spacecraft towards the Sun, the anomalous acceleration appears close to
$20 AU$ and up to $70 AU$, the anomalous acceleration seems to be a
constant with 10\% order of temporal and spatial variations of the
anomaly's magnitude. Turyshev {\it et al.}\cite{Turyshev2}
summarized recent results on researches of the anomaly.
Several conventional physical mechanisms have been proposed to
explain the anomaly, such as the unknown systematic--the gas leaks
from the propulsion system or a recoil force due to the on-board
thermal power inventory, and the conventional gravitational force
due to a known mass distribution in the outer solar system--the
Kuiper Belt Objects or dust, and the expansion of the universe
motivated by the numerical coincidence $a_P\simeq cH_0$. However, it was pointed out that
these conventional physical mechanisms can not be the answer of the
anomalous accelerations\cite{Anderson, Nieto}.

The failure of the conventional physical mechanisms imply that
the Pioneer anomaly may correspond to `new physics'. One of the most
popular `new physics' is the dark matter hypothesis. A specific
distribution of dark matter in the solar system would yield the wanted
result\cite{Foot}. However, this special distribution of dark matter
is not like the consequence of gravity. 
Thus, to explain the Pioneer anomaly, dark matter hypothesis still need more work. Several modified gravitational theories
also was suggested to explain the Pioneer anomaly, such as
the scalar-tensor vector gravity (STVG)theory\cite{Moffat}, brane-world models with large extra
dimensions\cite{Bertolami}, and conformal gravity with dynamic mass
generation\cite{Anderson}. These modified gravitational theories
seem appealing, however, most of them either much more complicated or
involves too much hypothesis which does not verified by experiments.

Finsler geometry, which takes Riemann geometry as its special case,
is a good candidate to solve the facing problems of the
theory of gravitation. The gravity in Finsler space has been studied
for a long time\cite{Takano,Ikeda,Tavakol, Bogoslovsky1}. In our
previous paper\cite{Finsler DM}, a modified Newton's gravity was
obtained as the weak field approximation of the Einstein's equation
in Finsler space of Berwald type. We have shown that the prediction
of the modified Newton's gravity is in good agreement with the
rotation curves of spiral galaxies without invoking dark matter
hypothesis.


Randers space, as a special kind of Finsler space, was first
proposed by G. Randers\cite{Randers}. Within the framework of
Finsler geometry, modified dispersion relation of free particle in Randers space has been
discussed\cite{RF}. A modified Friedmann model in Randers space is
proposed. It is showed that the accelerated expanding universe is
guaranteed by a constrained Randers-Finsler structure without
invoking dark energy\cite{Finsler DE}.

In this Letter, in the framework of Finsler geometry we will try to give a
simple and clear description of the Pioneer anomaly. As well-known, the length in
Riemann geometry is a function of positions. However, this is not the
case in Finsler geometry. In Finsler geometry, the length is a
function of both position and velocity. Finsler geometry is base on
the so called Finsler structure $F$ with the property
$F(x,\lambda y)=\lambda F(x,y)$, where $x$ represents position
and $y$ represents velocity. The Finsler metric is given as\cite{Book
by Bao}
 \be
 g_{\mu\nu}\equiv\frac{\partial}{\partial
y^\mu}\frac{\partial}{\partial y^\nu}\left(\frac{1}{2}F^2\right).
\ee

The Randers metric is a Finsler structure $F$ on $TM$ of the form
\be\label{Randers metric} F(x,y)\equiv\alpha(x,y)+\beta(x,y) ~,\ee
where \be \alpha(x,y)&\equiv&\sqrt{\tilde{a}_{\mu\nu}(x)y^\mu
y^\nu}\nonumber\\
\beta(x,y)&\equiv&\tilde{b}_\mu(x)y^\mu.\ee Here $\tilde{\alpha}$ is
a Riemannian metric on the manifold $M$. In this Letter, the indices
decorated with a tilde are lowered and raised by
$\tilde{\alpha}_{\mu\nu}$ and its inverse matrix
$\tilde{\alpha}^{\mu\nu}$, otherwise lower and raise the indices are
carried by $g_{\mu\nu}$ and $g^{\mu\nu}$. We will show that the
Finsler structure in Randers space with $\tilde{b}$ taking the
specific form $\tilde{b}_\mu=\{-kr,0,0,0\}$ corresponds
to the Pioneer anomaly. The above form of $\tilde{b}$ is given in
spherical coordinate and $k$ is a constant.

The parallel transport in Finsler space has been studied in terms of Cartan
connection\cite{Matsumoto,Antonelli,Szabo}. The notation of parallel
transport in Finsler manifold means that the length
$F\left(\frac{d\sigma}{d\tau}\right)$ is constant. Following the
calculus of variations, one gets the autoparallel equation in Finsler
space as\cite{Book by Bao} \be\label{geodesic of F}
\frac{d^2\sigma^\lambda}{d\tau^2}+\gamma^\lambda_{\mu\nu}\frac{d\sigma^\mu}{d\tau}\frac{d\sigma^\nu}{d\tau}=0.
\ee The autoparallel equation (\ref{geodesic of F}) is directly
derived from the integral length of $\sigma$ \be L=\int
F\left(\frac{d\sigma}{d\tau}\right)d\tau,\ee the inner product
$\left(\sqrt{g_{\mu\nu}\frac{d\sigma^\mu}{d\tau}\frac{d\sigma^\nu}{d\tau}}=F\left(\frac{d\sigma}{d\tau}\right)\right)$
of two parallel transported vectors is preserved. To get a modified
Newton's gravity, we consider a particle moving slowly in a weak
stationary gravitational field\cite{Weinberg}. Here, we suppose that
the Riemannian metric $\tilde{\alpha}$ is close to Minkowskian
metric, and $|\tilde{b}_\mu\tilde{b}^\mu|$ is very small
 \be \tilde{a}_{\mu\nu}(x)=\tilde{\eta}_{\mu\nu}+\tilde{h}_{\mu\nu}(x),
 \ee
where $\tilde{\eta}_{\mu\nu}$ is the Minkowskian metric and
$|\tilde{h}_{\mu\nu}|\ll1$.  Deducing from (\ref{geodesic of F}), we
obtain the geodesic of Randers space with constant Riemanian speed
(namely, $\alpha(\frac{d\sigma}{d\tau})$ is constant)
 \be\label{geodesic of R}
\frac{d^2\sigma^\lambda}{d\tau^2}+\tilde{\gamma}^\lambda_{\mu\nu}\frac{d\sigma^\mu}{d\tau}\frac{d\sigma^\nu}{d\tau}+\tilde{a}^{\lambda\mu}f_{\mu\nu}\alpha\left(\frac{d\sigma}{d\tau}\right)\frac{d\sigma^\nu}{d\tau}=0,
\ee where $f_{\mu\nu}\equiv\frac{\partial\tilde{b}_\mu}{\partial
x^\nu}-\frac{\partial\tilde{b}_\nu}{\partial x^\mu}$.

Randers\cite{Randers} has already found that the Randers metric is related to five dimensional Riemannian geometry. The five dimensional Riemannian metric $\gamma_{mn}$ ($m,n=1,2,3,4,5; \mu, \nu=1,2,3,4$) is given as
\begin{equation}
\gamma_{\mu\nu}=\tilde{a}_{\mu\nu}-\tilde{b}_\mu\tilde{b}_\nu;~\gamma_{\mu5}=\gamma_{5\mu}=\tilde{b}_\mu;~\gamma_{55}=-1.
\end{equation}
And the geodesic equation (\ref{geodesic of R}) of Randers metric is a solution of the five dimensional Einstein's field equation. The five dimensional Einstein tensor is expressed as
\begin{eqnarray}\label{field eq}
G^{\mu\nu}&=&\left(\tilde{R}^{\mu\nu}-\frac{1}{2}\tilde{a}^{\mu\nu}\tilde{R}\right)+\frac{1}{2}\tilde{E}^{\mu\nu},\\
G_5^{~\nu}&=&\frac{1}{2}f^{\nu\mu}_{~~~;\mu},\\
G_{55}&=&\frac{1}{2}\left(\tilde{R}-\frac{3}{4}f^{\mu\nu}f_{\mu\nu}\right),
\end{eqnarray}
where $\tilde{E}^{\mu\nu}=-f^{\mu\lambda}f_\lambda^{~\nu}+\frac{1}{4}\tilde{a}^{\mu\nu}f^{\lambda\theta}f_{\lambda\theta}$, $\tilde{R}^{\mu\nu}$ is the Ricci tensor of the four dimensional Riemannian metric $\alpha$, and the covariant derivative of four dimensional Riemannian metric $\alpha$ is denoted by ``;". In a geometrical viewpoint, the Randers metric arised from the Zermelo navigation problem \cite{Zermelo}. It aims to find the paths of shortest travel time in a Riemannian manifold under the influence of a drift (``wind"). Shen \cite{Shen} has shown that these minimum time trajectories are exactly the geodesics
of a particular Finsler geometry-Randers metric. The map between the Randers metric to a Riemannian space in the viewpoint of Zermelo navigation problem is investigated in the paper \cite{Gibbons}.

In weak field
approximation, the second term of the left side of the equation
(\ref{geodesic of R}) represents the Newtonian gravitational
acceleration. And the third term may induce the anomalous
acceleration. One should notice that the trajectory of the Pioneer 10
spacecraft is different from that of the Pioneer 11 spacecraft. The basic
property of the Pioneer anomaly that the direction of the anomalous
acceleration of the spacecraft towards the Sun tells us that the non vanish
components of $f_{\mu\nu}$ is $f_{0i}$. The term $f_{\mu\nu}$ acts
as electromagnetic force. In dealing with the
Pioneer anomaly, one need take only the ``electric force" into account.
Also, due to physical consideration, the ``electric force" should
be static. 
Thus, in the approximation of moving slowly and weak field, the
geodesic equation (\ref{geodesic of R}) reduces to
 \be\label{geodesic equation}
 \frac{d^2t}{d\tau^2}&=&0,\nonumber\\
 \frac{d^2\sigma^i}{d\tau^2}&=&-\frac{1}{2}\frac{\partial h_{00}}{\partial
 \sigma^i}\left(\frac{dt}{d\tau}\right)^2-\frac{\partial b_0}{\partial\sigma^i}\alpha\left(\frac{d\sigma}{d\tau}\right)\left(\frac{dt}{d\tau}\right).
 \ee
The solution of the first equation in (\ref{geodesic equation}) is
$dt/d\tau=const.$. Dividing the second equation in (\ref{geodesic
equation}) by $(dt/d\tau)^2$, we obtain
 \be
 \frac{d\sigma^i}{dt}=-\frac{1}{2}\frac{\partial h_{00}}{\partial
 \sigma^i}-\frac{\partial
 b_0}{\partial\sigma^i}.
 \ee
In spherical coordinate, the above equation changes as
 \be\label{acceler}
 a=\nabla\varphi+k,
 \ee where $\varphi\equiv-\frac{GM_\odot}{r}$ is the Newtonian gravitational potential. Then, from the equation
 (\ref{acceler}) one can see clearly that the anomalous acceleration
 \be
 a_p=k.
 \ee
Taking the average value of $a_p$, we can set the constant $k$
as $9.71\times10^{-25} m^{-1}$.

At a position of 20AU far from the Sun, the Newtonian
gravitational potential $\varphi=-4.43\times10^7 m^2/s^2$ and the
perturbation of Minkowskian metric $h_{00}=9.85\times10^{-10}$, and
$b_0=2.91\times10^{-14}$. Thus, the Finsler structure of Randers
space is a good description for metric fluctuation around the Minkowskian
one. At a position of 1AU far from the Sun, the
Newtonian gravitational potential $\varphi=-8.87\times10^8 m^2/s^2$
and the perturbation of Minkowskian metric
$h_{00}=1.97\times10^{-8}$, and $b_0=1.46\times10^{-15}$. Einstein's
relativity offers high order correction for Newtonian
mechanics (post-Newtonian approximation)\cite{Weinberg}, the
corresponding metric correction approximately equals $h_{00}^2$.
Here, we can see that $h_{00}^2$ is very
close to $b_0$. While the Pioneer is not far from the earth, it is
hard to distinguish the effect of general relativity(or Riemann
geometry) and Finsler geometry. This is a reason for why the anomaly
appears in the position of 20AU far from the Sun.

The equation (\ref{acceler}) implies that the modified gravitational potential is
\begin{equation}
\varphi_P=krc^2,
\end{equation}
where $c$ is the speed of light. Since the parameter $k$ is set as $9.71\times10^{-25} m^{-1}$, the ration $|\frac{\varphi_P}{\varphi}|$ is less than $10^{-8}$ for the solar system. The classical tests of general relativity are carried in solar system. Thus, the geodesic equation (\ref{geodesic of R}) also predict the same astrophysical phenomena that Einstein's
general relativity are able to predict. One also could directly obtain this fact from the field equation (\ref{field eq}), for the tensor $\tilde{E}^{\mu\nu}$ in it is the second order in $f^{\mu\nu}$.

The existence of the Pioneer anomaly suggests the Newton's theory of gravitation and general relativity need
to be modified even in the solar system. Here, we have suggested that Finsler
geometry could give a clear and simply description of the Pioneer
anomaly. The specific Finsler structure of the Randers space corresponds to the
Pioneer anomaly. We hope that the gravity anomalies mentioned in the
beginning of the Letter can be solved systematically in the
framework of Finsler geometry. 

\bigskip

\centerline{\large\bf Acknowledgements} \vspace{0.5cm}
 We would like to thank Prof. C. J. Zhu, H. Y. Guo and C. G. Huang for useful discussions. The
work was supported by the NSF of China under Grant No. 10525522 and 10875129.

\end{document}